\documentclass[preprint,1p,authoryear]{elsarticle}

\PassOptionsToPackage{hyphens}{url}\usepackage[colorlinks=true, linkcolor=blue, citecolor=blue, urlcolor=blue]{hyperref}

\usepackage{url}
\usepackage{amsmath}
\usepackage{graphicx}
\usepackage{booktabs}
\usepackage{multirow}
\usepackage{tabularx}
\usepackage{balance}
\usepackage{booktabs}
\usepackage{placeins}
\usepackage{threeparttable}

\begin{document}

\title{A Text Mining Analysis of Data Protection Politics: \\ The Case of
  Plenary Sessions of the European Parliament}

\author[utu]{Jukka Ruohonen\corref{cor}}
\ead{juanruo@utu.fi}
\address[utu]{Department of Computing, University of Turku, FI-20014 Turun yliopisto, Finland}
\cortext[cor]{Corresponding author.}

\begin{abstract}
Data protection laws and policies have been studied extensively in recent years,
but little is known about the parliamentary politics of data protection. This
limitation applies even to the European Union (EU) that has taken the global
lead in data protection and privacy regulation. For patching this notable gap in
existing research, this paper explores the data protection questions raised by
the Members of the European Parliament (MEPs) in the Parliament's plenary
sessions and the answers given to these by the European Commission. Over a
thousand of such questions and answers are covered in a period from 1995 to
early 2023. Given computational analysis based on text mining, the results
indicate that (a)~data protection has been actively debated in the Parliament
during the past twenty years. No noticeable longitudinal trends are present; the
debates have been relatively constant. As could be expected, (b) the specific
data protection laws in the EU have frequently been referenced in these debates,
which (c) do not seem to align along conventional political dimensions such as
the left-right axis. Furthermore, (d) numerous distinct data protection topics
have been debated by the parliamentarians, indicating that data protection
politics in the EU go well-beyond the recently enacted regulations.
\end{abstract}

\begin{keyword}
data protection, privacy, parliamentarism, agenda-setting, text mining, topic
modeling, EU, GDPR
\end{keyword}

\maketitle

\section{Introduction}

Data protection has a long history in Europe. There were national data
protection laws in many European countries long before the General Data
Protection Regulation (GDPR) repealed an earlier 1990s data protection directive
in the EU in 2016. The early efforts at the European level started already in
the late 1960s with initiatives from the Council of Europe, which, however, did
not progress despite a positive reception at the European Parliament during the
1970s~\citep{Evans81}. In the 1980s the data protection initiatives switched
from the European Community to other institutions, including the Organisation
for Economic Co-operation and Development (OECD) and the United Nations. Things
started to change when the EU was formally established with the Maastricht
Treaty in 1992. During the policy-making of the Lisbon Treaty in the late 2000s,
data protection was added to all basic treaties of the EU, including the Treaty
on European Union, the Treaty on the Functioning of the European Union, and the
Charter of Fundamental Rights of the European Union. In addition, privacy (but
not data protection) has long been present in the European Convention on Human
Rights. The GDPR is the latest manifestation in this long historical trajectory
in Europe.

The European Commission is the guardian of these treaties and the fundamental
rights specified in them. According to anecdotal evidence, however, it has often
been the European Parliament who has taken the political initiative and lead in
promoting data protection and privacy policies, which have often received less
attention in national parliaments~\citep{Mistale15}. A~comparable point applies
regarding the enforcement of the data protection policies in the European
Union. Here, the final say in many difficult data protection questions has often
been done by the Court of Justice of the European Union (CJEU), although
enforcement of data protection is generally delegated to national data
protection authorities in the member states. According to a critical viewpoint,
neither the guardian of the treaties nor these data protection authorities have
fulfilled their data protection obligations to an extent envisioned in the
enacted treaties and laws.

But law is one thing and lawmaking is another thing. Regarding the latter, only
little is generally known about the question of how politicians perceive data
protection and take it into account in their political agenda-setting. True
enough, there have long been European politicians who have actively profiled
themselves as promoters of data protection and privacy, but the large majority
of politicians has supposedly perceived these issues as having a lesser weight,
questions related to national security and related topics perhaps
notwithstanding. In this regard, the intense politics during the policy-making
of the GDPR, as adeptly described in a film documentary~\citep{Powles15}, were
arguably an exception rather than the rule in the sense that the political data
protection issues and the inner-workings of the Parliament became visible also
to the general public in Europe and elsewhere. Against this background, on the
one hand, it could be argued that data protection politics often deal with
obscure technical details and boring nitty-gritties in law proposals, which
seldom prompt a wider political interest among the constituencies of
politicians. On the other hand, it could be argued that the continuous stream of
privacy scandals, whether in terms of data breaches or the practices of social
media companies, the increased state surveillance, including the associated
oppression and human rights violations in authoritarian regimes, the rise of
large technology companies and new technologies such as facial recognition, and
many related issues would provide valuable political capital for politicians in
their recurrent seeking of votes and offices.

The European Commission is the main lawmaking institution in the European
Union. It makes legislative proposals. If a proposal is then agreed by the
Council of the European Union and European Parliament, a new law is
enacted. Although this basic lawmaking procedure has not changed, the Parliament
has gained power as a budgetary authority and a co-legislator with the Lisbon
Treaty, although it has at the same time faced the rise of eurosceptic and
populist parties in the aftermaths of the various crises since 2008. The
Parliament possesses also considerable power as a political agenda-setter in the
pre-legislative phase; it can influence the legislative agenda indirectly by
bringing new issues into the EU's political roundtables and further escalating
these to the fore of public debates~\citep{Kreppel19}. Such political
agenda-setting frames the context of the present work: the goal is to explore
data protection questions raised by MEPs in the Parliament's plenary sessions
and the answers given to these by the Commission. If data protection and privacy
more generally provide political capital for MEPs as envisioned, there should
be also many associated questions about these topics. The exploration can
further reveal important aspects about those specific data protection issues
that politicians perceive as important. As further argued in the opening
Section~\ref{sec: background}, the exploration has novelty because there are no
directly comparable previous works and the knowledge about data protection
politics is generally very limited. Then, the materials and methods are
elaborated in Section~\ref{sec: materials and methods}. The results from the
empirical exploration are presented in Section~\ref{sec: results}. The final
Section~\ref{sec: conclusion} concludes.

\section{Background}\label{sec: background}

\subsection{Data Protection Politics}

The national political debates around data protection and privacy vary across
Europe. In some countries, such as Ireland and Romania, there has been hardly
any debates at all, while moderate political debates have been present in other
countries such as France, and more intense debates in countries such as the
Netherlands, Germany, Sweden, the United Kingdom, and
Italy~\citep{Custers18}. As is the case throughout the world, a typical dividing
line in many European countries has been between privacy and national
security. Human rights have been emphasized in some countries more than in
others. A further persistent issue has been the relation between data protection
and economic issues, including the competitiveness of smaller European companies
against large multinational technology companies. Also attention given to data
protection issues by media varies across Europe, as does public opinion on these
issues. However, little is still known about parliamentary politics regarding
data protection even in the context of the European Union within which much of
the legislative data protection work occurs.

In general, the existing research is scarce on data protection
\textit{politics}, while, at the same time, there is an abundance of legal
scholarship on data protection laws and \textit{policies} as well as on the
impacts of these, whether upon technical systems studied in computer science or
the scientific research practices in medicine. This situation is surprising
because data protection, at least from a European perspective, is vital for the
preservation and promotion of political participation, political systems, and
democracy through its provisions for many fundamental rights \citep{Naef23}. The
occasional intensity in both research and politics seems to have spiked around
distinct events, the most well-known of such events being the surveillance
machinery that was revealed by Edward Snowden.

Also other controversies have occasionally been addressed in research from the
perspective of data protection politics. A good example would be the political
cleavages that were initially present regarding the data retention Directive
2006/24/EC. For various reasons, the Parliament made a U-turn during the
policy-making of this directive, switching its data protection rationale to one
based on security~\citep{Servent13}. To some degree, the earlier position was
justified, given that the CJEU later invalidated the directive in 2014. Another
good example would be the so-called passenger name record (PNR) debacle in the
early 2010s, which led to the enactment of the corresponding Directive (EU)
2016/681. During the policy-making of this directive, which overlapped with
questions related to the SWIFT system and the so-called Safe Harbour agreement
that was invalidated by the CJEU in 2015, the Council, the Commission, and the
Parliament clashed many times, mainly due to the Parliament's critical stance on
data protection violations~\citep{Huijboom15, Mistale15,
  Servent15}.\footnote{~The Society for Worldwide Interbank Financial
  Telecommunication (SWIFT).} In this case and generally during the 2010s, the
green and left-wing (but not the social democratic) political groups in the
Parliament mostly voted against legislative proposals they perceived as
violating data protection laws and fundamental rights, the votes cast by the
liberal group split in half, and the conservative and social democratic groups
voted in favor~~\citep{Servent15}. Similar but not entirely uniform results
apply to the politics of making the GDPR. Given the immense lobbying from both
industry associations and civil society groups, the green, left-wing, and social
democratic groups submitted amendments that strengthened the GDPR's data
protection provisions, while the submissions from the liberal, centre-right, and
eurosceptic groups predominantly weakened these~\citep{Christou21,
  Hilden19}. These few examples notwithstanding, a fairly comprehensive
literature search indicates little---if any---further valuable works on the
political aspects of data protection. Therefore, the paper's text mining
approach to the Parliament's plenary sessions patches an important gap in
existing research---even when keeping in mind the exploratory research design
that necessarily comes with quantitative text mining. That is to say, explicit
hypotheses are neither presented nor confirmed; the goal is to quantitatively
explore the topic at hand. The exploratory approach is justifiable due to the
lack of existing research that could be used to guide a confirmatory research
approach.

\subsection{Plenaries}

The plenary sessions of the European Parliament take place every month in
Strasbourg with additional meetings held in Brussels. These carry an important
political function for MEPs and their constituencies in Europe and in the member
states. Nowadays, the sessions are even streamed online. Typically, either MEPs
raise oral and written questions targeted toward the Council and particularly
the Commission or they respond to an opening statement from the Commission,
followed by a response by a rapporteur from a given committee in the
Parliament. Debate follows. As in most national parliaments, time allocated for
speaking is restricted. After the debate is over, the Commission concludes and
notes its position on any proposed amendments to legislative proposals.

Yet, the function of the plenaries is most of all political: these seldom go
into legislative details such as amendments, which are primarily handled in
committees within which also the ``real discussions'' typically
occur~\citep{Lord13, Servent15}. From an institutional viewpoint, the plenaries
mainly publicly justify the Parliament's positions and decisions on given policy
matters. For individual MEPs, on the other hand, these provide publicity and a
venue for outlining their ideological positions and political
viewpoints.

Regarding these ideologies, it has been observed that the positions taken by
MEPs in their plenary speeches do not align with the usual left-right axis, but
instead tend to reflect pro-Europe versus eurosceptic viewpoints and divisions
in national politics in the member states~\citep{Proksch10}. The same applies to
the volume of speeches for which national party ideology but not the left-right
axis at the EU-level has been observed to matter~\citep{Greene15}. Though, the
situation is complicated because the left-right axis at the European Parliament
does not necessarily reflect similar axes in the member states. The political
positions taken by MEPs tend to also vary across policy domains. The traditional
left-right dimension is still important in the Parliament regarding economic and
market policies as well as those related to redistribution and labor
markets~\citep{Borzel22}. In some other highly contested policy areas such as
asylum, the positions taken by MEPs have also been observed to align with their
national parties' left-right preferences and not their stances on European
integration~\citep{FridNielsen18}. There are also various other issues at
play. Even practicalities matter. For instance, frontbenchers (who sit in front
of their group) have been observed to deliver more questions and speeches than
backbenchers~\citep{Sorace18}, and so forth. Against this brief take on existing
empirical results, it is difficult to deduce clear-cut hypotheses for the
ideological background and potentially existing political cleavages behind
questions raised about data protection in the plenaries.


\subsection{Related Work}

There are two branches of related work. Both have already been briefly
addressed; these are the political aspects of data protection and the plenary
sessions of the European Parliament. Given the paper's exploratory approach, it
is worthwhile to pinpoint a few details about the methodological approaches
taken in the latter research branch.

While qualitative methods have sometimes been used to codify the contents of the
plenary debates~\citep{Lord13}, most of the recent empirical research seems to
generally rely on quantitative methods. These methods range from social network
analysis~\citep{Walter22} to regression analysis \citep{Sorace18}. Closer to the
present work, there are also various previous studies that have used text mining
techniques to analyze the plenaries. Among other things, techniques based on
word counts have been used~\citep{FridNielsen18, Wei20}. Also specific methods
for parsing and pre-processing have been presented~\citep{Aggelen17}. Even
closer to the present work, topic modeling has been used to analyze the contents
of the plenary debates~\citep{Greene15}. Based on this brief review, it can be
concluded that the paper's exploratory approach is nothing special as such; also
many of the previous works are about exploration rather than confirmation. Nor
is there novelty in the methodological approach taken. But what is novel is the
focus on data protection politics. No previous works exist in this regard.

\section{Materials and Methods}\label{sec: materials and methods}

\subsection{Data}

The empirical dataset contains questions raised by MEPs about data protection in
the plenary sessions of a given European Parliament and the answers to these by
a given European Commission. These were retrieved from the Parliament's online
database on 7~February 2023 by using the two words ``data protection'' for the
keyword
search.\footnote{~\url{https://www.europarl.europa.eu/plenary/en/parliamentary-questions.html}}
Both oral and written questions by parliamentarians are covered, while the
answers from a given Commission are always delivered in writing. The period
covered includes the Commissions of Jacques Santer (1995 -- 1999), Romano Prodi
(1999 -- 2004), Jos\'e Manuel Barroso (2004 -- 2014), Jean-Claude Juncker (2014
-- 2019), and Ursula von der Leyen (from 2019 to the present day). As not all
questions have received answers, the dataset contains $1230$ questions and
$1163$ answers. These were merged together and thus the sample size is $n =
1230$. Henceforth, the term document is used to refer to the answer-question
pairs. Despite the merging, however, the longitudinal reference points are the
dates on which the questions were raised by the parliamentarians in the
plenaries and not the dates upon which the answers were received to the
questions asked.

\subsection{Processing}

The archival material on the questions raised in the plenary sessions is mostly
unstructured, free-form text. Each question reflects the individual writing and
argumentation style used by a particular parliamentarian or a group of
parliamentarians. In other words, these are not legal documents with a
well-defined syntax and a common vocabulary. Neither is there an existing corpus
that could be used for quantification. By implication, existing techniques
developed for empirical legal analysis cannot be readily applied (see
\citealt{Ruohonen22IS} for a discussion about these techniques). A more generic,
semantics-agnostic approach was therefore used for pre-processing the archival
material.

Many of the questions trace to topical data protection issues, controversies,
and scandals, whether national, European, or global. Many short and even na\"ive
questions have been asked about data protection, often based on articles in
newspapers and their scandalous headlines. Such questions reflect the nature of
the Parliament's plenary sessions: as said, these seldom deal with actual
legislative work---the function of these is mainly political. Nevertheless, the
dataset contains also some specific and well-researched questions that have
addressed some particular laws or proposals thereto, articles in these, and even
their recitals. The sample applies to the Commissions' replies that have usually
been stylish and structured, usually containing references to specific
regulations, directives, decisions of a given Council or a Commission, and a
given Commission's legislative proposals and communications (COMs) for a given
European Parliament and a Council. These structural elements were exploited for
quantifying information with regular expression searches on directives,
regulations, decisions, and proposals, using the common syntactic naming
conventions in the EU's jurisprudence, policy-making, and document archiving. In
what follows, these are together referred to as policies for brevity.

In addition, regular expression searches were used to parse the political
affiliations of the MEPs raising the questions. To simplify the analysis, all
historical political groups were merged to align with the current composition of
the Parliament, hard and soft eurosceptic groups were unified, and a group of
others was reserved for the rest. Thus, the analysis operates with seven groups:
a centre-right group (including the current European People's Party), a liberal
group (including the current Renew Europe), a social democratic group (including
the current Progressive Alliance of Socialists and Democrats), a left-wing group
(including the current Left in the European Parliament), a green group
(including the current Greens/European Free Alliance), eurosceptics, and
others.\footnote{~To be more specific: the centre-right group includes the
  current EPP and the historical UPP, the liberal group includes the current
  Renew as well as the historical ALDE and ELDR, the social democratic group
  includes the current S\&D and the historical PSE, the group of eurosceptics
  contains the current ID and ECR groups as well as the historical ITS, EFD,
  IND/DEM, ENF, UEN, and EFDD groups, and the others include the current NI
  group and the historical TGI group. Also French abbreviations were used for
  parsing. The abbreviations themselves can be looked up from Wikipedia, among
  other sources.} Although this merging is not perfect, it suffices for a
quantitative analysis because all of the main ideological positions are still
represented in the seven-fold classification used. It should be further remarked
that many questions were asked by informal groups of MEPs with possibly
different affiliations. In these cases, all of the affiliations were counted;
therefore, the number of affiliations ($1643$ in total) exceeds the number of
questions.

Then, a bag-of-words corpus was constructed from the documents. This
construction involved nine steps: (1) all hypertext markup language elements
were removed; (2) all textual data was lower-cased; (3) the lower-cased text
data was tokenized according to white space and punctuation characters; (4) only
alphabetical tokens were included; (5) stopwords were omitted; (6) tokens with
lengths less than three or more than twenty characters were excluded; (7) two
adjacent tokens were used for constructing bigrams; (8) term-frequencies were
calculated for the bigrams; and (9) those bigrams were excluded that occurred in
the corpus less than three times. The \citet{NLTK19} package was used for the
pre-processing.\footnote{~In addition to the standard stopwords in the toolkit,
  the following custom stopwords were used: \textit{question}, \textit{answer},
  \textit{debate}, \textit{data}, \textit{protection}, \textit{legal},
  \textit{policy}, \textit{text}, \textit{texts}, \textit{rule}, \textit{rules},
  \textit{personal}, \textit{persons}, \textit{individuals}, \textit{natural},
  \textit{processing}, \textit{european}, \textit{union}, \textit{member},
  \textit{state}, \textit{states}, \textit{parliament}, \textit{council},
  \textit{commission}, \textit{directive}, \textit{regulation},
  \textit{decision}, \textit{com}, \textit{final}, \textit{article},
  \textit{tabled}, \textit{reply}, \textit{deadline}, \textit{within},
  \textit{update}, \textit{updated}, \textit{regard}, \textit{behalf},
  \textit{forward}, \textit{forwarded}, \textit{aware}, \textit{given},
  \textit{consider}, \textit{intend}, \textit{notice}, \textit{take},
  \textit{taking}, \textit{took}, \textit{place}, \textit{account},
  \textit{would}, \textit{could}, \textit{like}, \textit{last},
  \textit{concerning}, \textit{ensure}, \textit{inter}, \textit{alia},
  \textit{agree}, \textit{http}, \textit{https}, and all calendar time months.}
The only noteworthy remark about these rather conventional pre-processing steps
relates to bigrams (or $2$-grams); these are used because of improved
interpretability. Due to the ninth pre-processing step, the amount of documents
in the corpus reduced to $1220$. The total amount of bigrams is $13366$.

\subsection{Methods}

Descriptive statistics and topic modeling are used for the empirical
analysis. In terms of the latter, the well-known and well-documented latent
Dirichlet allocation (LDA) method is used~\citep{Blei12, WangBlei11}. In
essence, this Bayesian method extracts a finite number of latent topics from a
corpus. Each topic is a mixture of bigrams with different proportions and each
document is a mixture of latent topics, again with different proportions, such
that all documents in the corpus share the same set of topics but different
documents exhibit the topics in different portion.

In applied research the perhaps biggest issues in using LDA is the difficult
choice over the number of topics to extract. To help in making the choice, the
algorithms developed by \citet{CaoXia09}, \citet{Devaud14}, and
\citet{Griffiths04} are used, as implemented in an R
package~\citep{ldatuning20}. The first algorithm is about minimizing an
objective function, while the latter two are about maximization; hence, smaller
and larger values are better. For computation, the Gibbs sampler was specified
for all three algorithms, which were then executed for $2, 3, \ldots, 100$
number of topics. After having chosen the number of topics, the final
computation is then carried out with a related R package~\citep{topicmodels},
again using the Gibbs sampler. As there are appears to be neither computational
methods nor heuristic guidelines for choosing the two associated hyperparameters
(usually denoted as $\alpha$ and $\beta$) for the LDA
method~\citep{Nikolenko17}, the R package used was allowed to estimate these.

After having the final LDA model computed, the dominant topic for each
document (that is, the most likely topic in a document) is linked to the
political group of a particular MEP or political groups in case multiple MEPs
were involved in asking the corresponding question. These links are used as
edges in an undirected but weighted (social) network; the weights represent the
times a MEP from a given political group asked a question with a given dominant
topic. The vertices refer to the topics and the political groups. The resulting
network resembles the co-word networks that have been used to analyze the
plenary sessions~\citep{Wei20}, although topics are used instead of words.

\section{Results}\label{sec: results}

\subsection{Policies}

\begin{figure}[th!b]
\centering
\includegraphics[width=\linewidth, height=6cm]{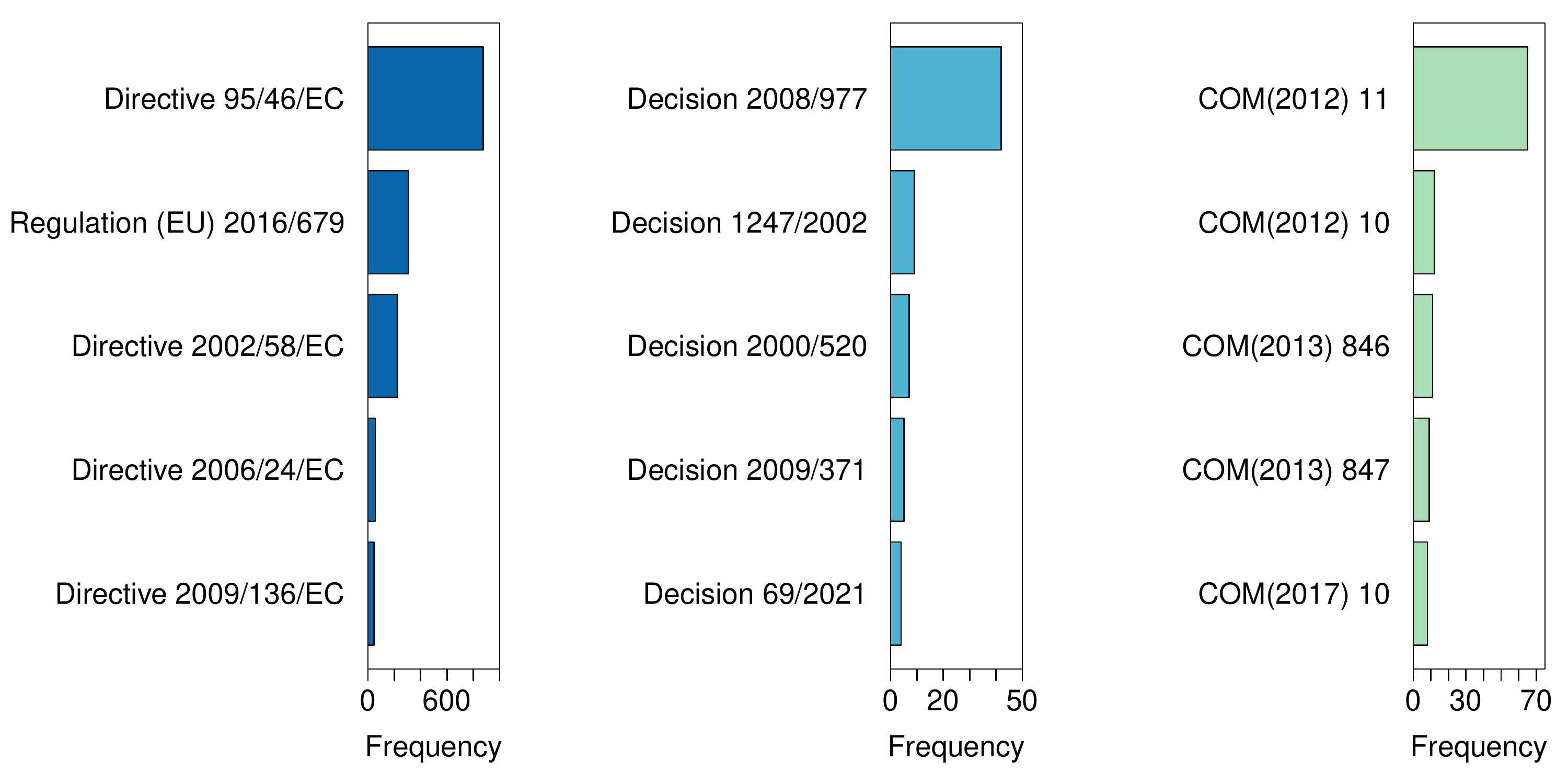}
\caption{The Top-5 Directives, Regulations, Decisions, and Communications
  Referenced}
\label{fig: laws}
%
\vspace{15pt}
%
\centering
\includegraphics[width=\linewidth, height=3cm]{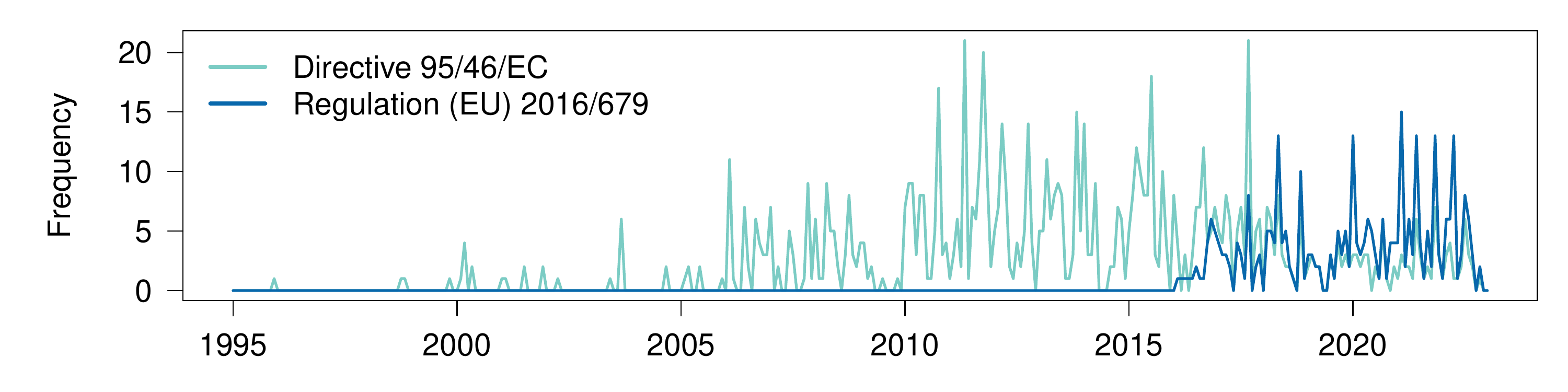}
\caption{References Made to the GDPR and Directive 95/46/EC}
\label{fig: gdpr}
\end{figure}

The regular expression searches outputted $155$ references to regulations and
directives, $46$ references to decisions, and $166$ references to communications
and proposals. Fig.~\ref{fig: laws} displays the top-5 rankings for these. The
most frequently referenced law in the documents has been the legacy data
protection Directive 95/46/EC, which was repealed by the GDPR, that is,
Regulation (EU) 2016/679. This result is hardly surprising as such because the
legacy directive was in force over twenty years before finally having been
replaced by the Juncker Commission. As can be further seen from the time series
illustration in Fig.~\ref{fig: gdpr}, occasional references to the legacy
directive started to first appear in the early 2000s. These then increased until
the enactment of the GDPR in 2016. Thereafter, the GDPR has stolen the
referencing, although, interestingly enough, the old directive was sometimes
referenced also during the late 2010s and early 2020s. To some extent and
implicitly, the increased volatility visible in the reference counts after
around 2010 supports the commonly raised argument \citep[see, e.g.,][]{Rossi18}
that the Snowden revelations in 2013 had an important impact upon the
policy-making of the GDPR.

Also the so-called e-privacy Directive 2002/58/EC has received a considerable
amount of references. This observation too is understandable because the
e-privacy directive is the current EU-level law for regulating important privacy
aspects in telecommunications, such as the confidentiality of information and
the treatment of traffic data, as well as other related issues, such as web
cookies and even spam. Reforming the directive has also long been on the
political agenda in the EU, but thus far to no avail. As for the decisions,
Decision 2008/977 has clearly been the most referenced one. Once again, the
observation is understandable: it was a decision addressing data protection in
law enforcement and criminal justice matters until its replacement by Directive
(EU) 2016/680, which was enacted together with the GDPR. Also references to the
Commissions' proposals align with these results; COM(2012) 11 was the
proposal upon which the GDPR was legislated.

\subsection{Political Groups}

\begin{figure}[th!b]
\centering
\includegraphics[width=\linewidth, height=12cm]{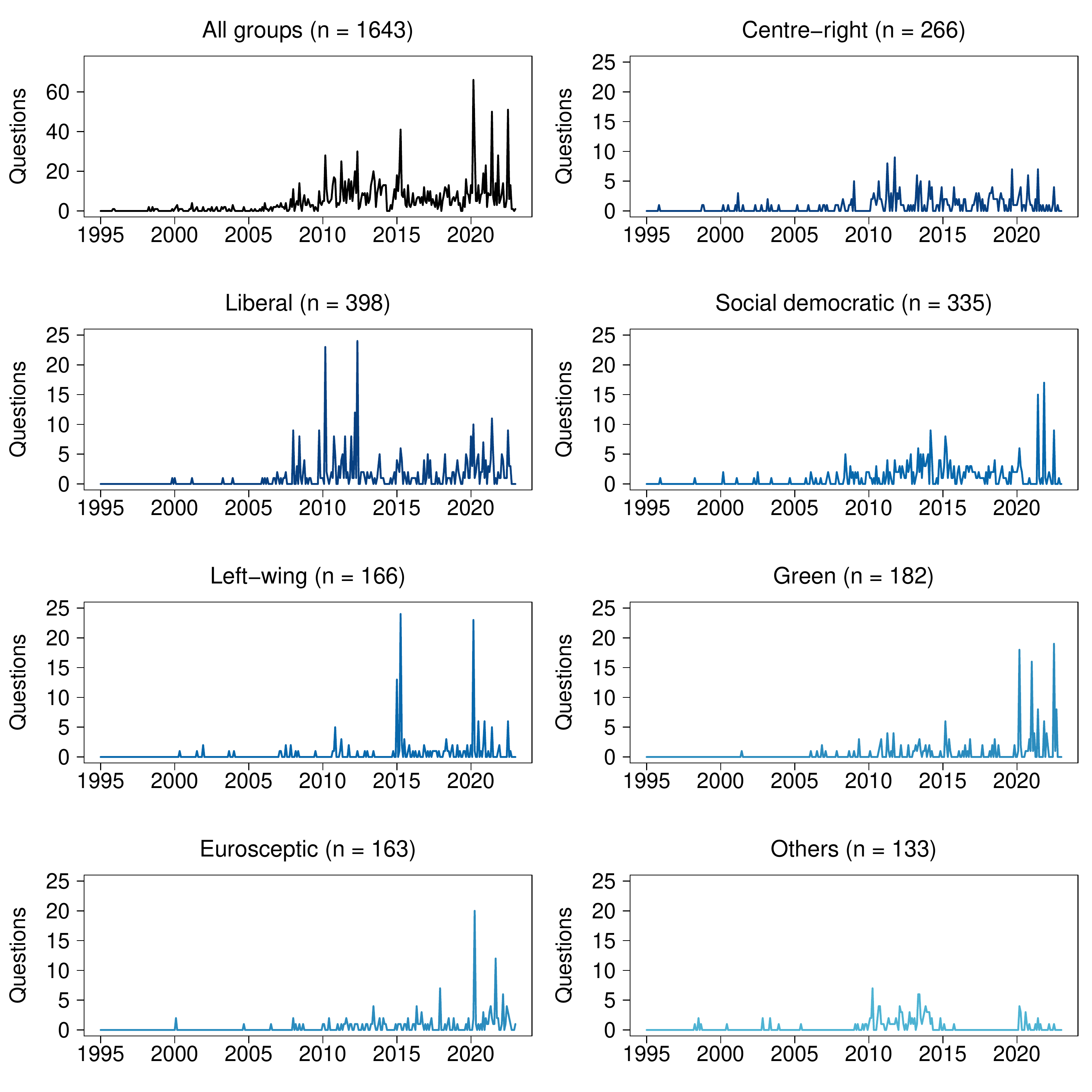}
\caption{Political Affiliations of the MEPs Involved in the Questions}
\label{fig: groups}
%
\vspace{15pt}
%
\centering
\includegraphics[width=\linewidth, height=5cm]{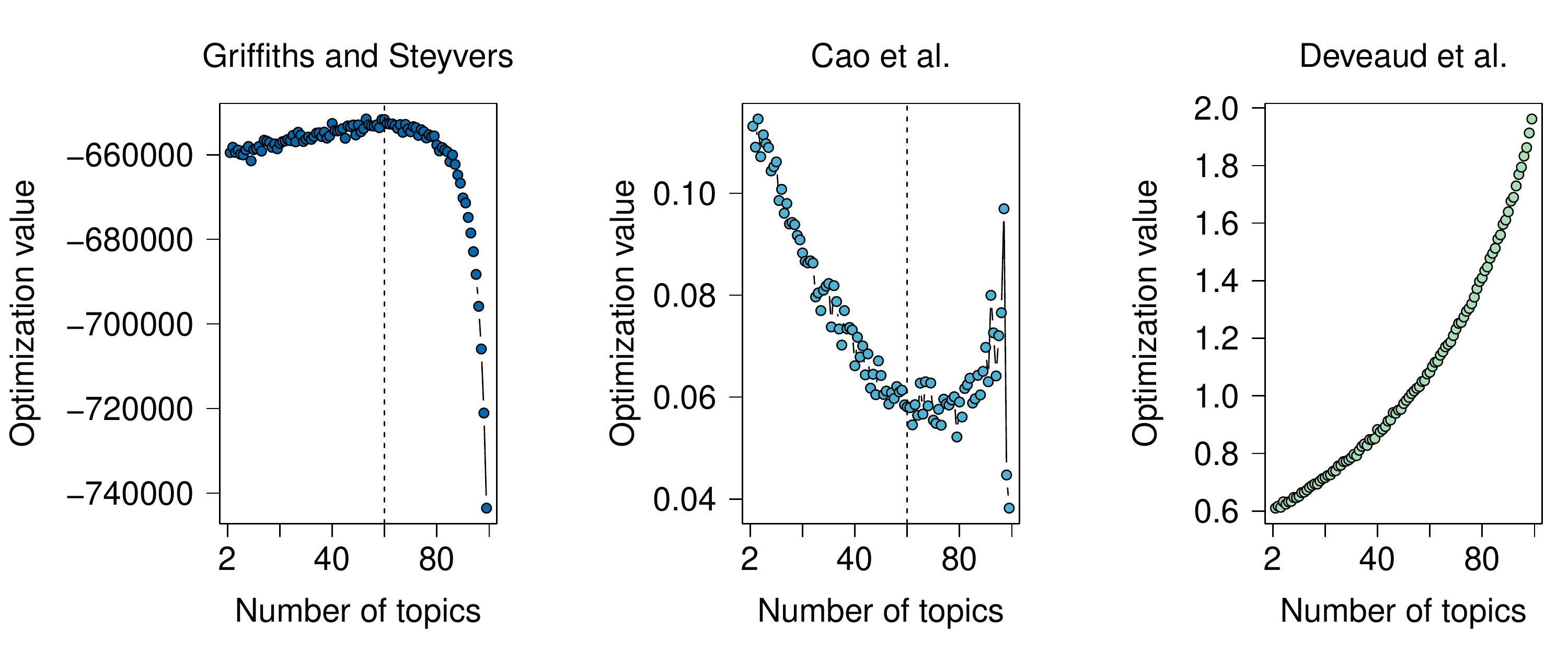}
\caption{Algorithmic Tuning Results for the Number of Topics}
\label{fig: lda tuning}
\end{figure}

The volume of data protection questions raised by MEPs belonging to specific
political groups can reveal insights about whether there are noteworthy
ideological divisions. To this end, Fig.~\ref{fig: groups} shows the number of
questions raised according to the seven political groups together with a time
series of all questions raised. The volume of all data protection questions
asked started to increase in the early 2010s. But unlike the earlier time series
for references made to Directive 95/46/EC and the GDPR in Fig.~\ref{fig: gdpr},
the volume of all questions asked indicates intensification in the early
2020s. In particular, there have been three noteworthy spikes in recent
years. These spikes are addressable to MEPs belonging particularly to the social
democratic, left-wing, green, and eurosceptic groups.

Parliamentarians belonging to the liberal group have raised the largest amount
of data protection questions in overall, but these occurred particularly during
the first half of the 2010s. Interestingly, furthermore, MEPs belonging to the
centre-right group, which includes EPP that has been the largest group in the
Parliament since 1999, have shown only modest interest in data protection. A
further point is that left-wing MEPs raised many questions around 2015 during
which the GDPR was debated---even though the rapporteur for the regulation was
from the green group. In general, the European greens seem to have started to
show political interest in data protection only later on in the 2020s, at least
when judged by the volume of questions raised by their MEPs. All in all, it
seems that data protection has been debated in the plenaries particularly by
parliamentarians representing liberal and social democratic groups. Those
representing left-wing, green, and eurosceptic groups have joined in more recent
times.

\subsection{Topics}

The topic modeling exposition can be started by examining the outputs from the
noted algorithms for estimating the number of topics automatically. These are
shown in Fig.~\ref{fig: lda tuning}. The third algorithm is clearly
uninformative, while the first algorithm seems to indicate an optimum around
eighty topics and the second algorithm around sixty topics or so. When
estimating LDA models with the number of topics in the range \text{60 -- 80},
many sharp, consistent, and interpretable topics emerge, but, at the same time,
several topics cannot be readily interpreted and labeled. A potential
explanation for these uninterpretable topics relates to the noted stylistic
conventions used by the MEPs in their plenary questions; the interpretability of
LDA models depends not only on pre-processing but also on lexical and stylistic
homogeneity of a given corpus~\citep{Sherstinova22}. In any case, these
preliminary computations reveal the inherently difficult choice over the number
of topics. One possibility would be to estimate a model with around 60 topics
and then only focus on those that can be interpreted, but such an approach would
be open to criticism about cherry-picking~\citep{Brookes19}. Another possibility
would be to prefer the ``humanistic side'' with its emphasis on
interpretability~\citep{Paakkonen21} and estimate a model with fewer topics, but
this choice would lead to generally incoherent topics with many overlapping
bigrams. To strike a middle ground, the final model was estimated with 24
topics. The top-10 bigrams with their posterior probabilities for these topics
are shown in Appendix; see Figures~\ref{fig: lda bigrams 1} and \ref{fig: lda
  bigrams 2}.

All of the twenty-four topics could be reasonably interpreted and
labeled. However, there are three cases that are represented by two topics each:
telecommunications, the GDPR, and the PNR. These were merged together for
further analysis. In addition, there is one topic that does not represent data
protection but rather gathers bigrams related to practicalities of the plenary
sessions, such as procedures, documents, and voting. Otherwise, the interpreted
topics reveal interesting aspects about data protection and privacy issues
raised by MEPs in the parliamentary plenary sessions. Fourteen points can be
made to briefly elaborate these issues and the corresponding topics, as follows:

\begin{enumerate}

\itemsep 5pt

\item{The enduring big questions related to transatlantic data flows are
  well-represented. As already noted, the CJEU annuled the Safe Harbor agreement
  between the United States and the EU in 2015. Later on, in 2020, the court
  further invalidated the agreement's successor, that is, the so-called Privacy
  Shield agreement. Both agreements attain their own topics. Also the CJEU
  itself has its own topic.}

\item{Data protection questions related to immigration and free movement within
  the EU are represented. In terms of the latter, there is a distinct topic for
  bigrams related to the Schengen area. The immigration question is more
  interesting: it is primarily represented by the European Dactyloscopy
  (Eurodac) database, which is used to identify asylum seekers and irregular
  border-crossers based on their fingerprints, pictures, and other
  biometrics. This kind of data belongs to the GDPR's category of sensitive
  data, and, therefore, it is no surprise that also MEPs have raised data
  protection questions about the Eurodac database. It has received data
  protection scrutiny and associated criticism also in academic
  research~\citep{Boehm12}.}

\item{Early on during the COVID-19 pandemic, there was a big debacle about the
  mobile applications for contact tracing. In Europe this debacle centered
  around privacy questions and the implementations of large technology companies
  for the tracing functionality \citep{Tretter23}. Also this debacle is
  represented with its own topic.}

\item{As could be expected, there is a topic for social networks and social
  media. It has also gathered questions related to facial recognition, probably
  due to the Clearview AI company that gathers images from social media and the
  Internet in general.}

\item{Documents about intellectual property and property rights are represented
  in their own distinct topic, although there is some incoherence present, as
  also such bigrams as \textit{public health}, \textit{medical products}, and
  \textit{health research} appear in the top-10 ranking. This observation
  reflects the choice to prefer 24 topics; with a higher number of topics, the
  medical and healthcare documents would likely attain their own topic.}

\item{There is a topic for documents related to law enforcement and criminal
  justice. In addition to the already noted data protection Directive (EU)
  2016/680 for law enforcement and criminal matters, it is worth remarking
  Directive 2014/41/EU for cross-border criminal investigations in the EU. Both
  contain their own data protection and fundamental rights issues
  \citep{Jesserand18, Tosza20}.}

\item{Documents about national authorities and supervisory bodies belong to
  their own topic. Given the enduring problems with the GDPR's enforcement by
  national data protection authorities \citep{Ruohonen22IS}, this result is not
  surprising.}

\item{Aviation is the label used for one topic. This outlying topic is somewhat
  incoherent; it is represented by such bigrams as \textit{judicial
    cooperation}, \textit{secure flight}, \textit{third countries},
  \textit{police judicial}, and \textit{homeland security}. Thus, some overlap
  is present with the topic on law enforcement and those dealing with
  transatlantic data flows.}

\item{Cloud computing involves many difficult data protection
  questions~\citep{Celeste21}. Therefore, it is understandable that there is a
  specific topic for it.}

\item{There is a topic labeled according to the top-ranking bigram \textit{body
    scanners}. Upon a brief qualitative look at the actual questions involving
  this bigram, these body scanners refer to those used in airports. Many data
  protection, privacy, and human rights questions were raised about these
  scanners in the late 2000s and early 2010s.}

\item{The single market of the European Union is also present with its own
  topic. Although some incoherence is present also in this topic, data
  protection questions and answers related to the single market are
  understandable already because the single market is perhaps the most
  significant element of the EU as a whole. The Parliament has been also active
  as a co-legislator in the various recent reforms and proposals for a renewed
  digital single market powered by artificial intelligence, data, and other
  related aspects (for the background see, e.g., \citealt{JustoHanani22}).}

\item{The INDECT research project has its own topic. This EU-funded project was
  about providing intelligent security systems for law enforcement agencies
  \citep{Stoianov15}. Analogously to many related large-scale EU-funded security
  research programs, it was subjected to concerns about privacy, surveillance,
  and ethics.}

\item{There is a topic about Hungary. This topic is related to the
  persistent rule of law issues and controversies regarding this member
  state. Apparently, these issues involve also data protection violations by
  Hungarian authorities according to MEPs.}

\item{Last but not least, human rights and fundamental rights belong to their
  own distinct topic. As noted, these are the foundation for data protection and
  privacy in Europe.}

\end{enumerate}

Thus, a myriad of different data protection questions have been asked and
debated in the plenary sessions of the European Parliament. Overall, these topic
modeling results underline that data protection touches various issues in the
European Union---even though the GDPR has stolen much of the attention in both
research and practice.

\begin{figure}[th!b]
\centering \includegraphics[width=\linewidth,
  height=12cm]{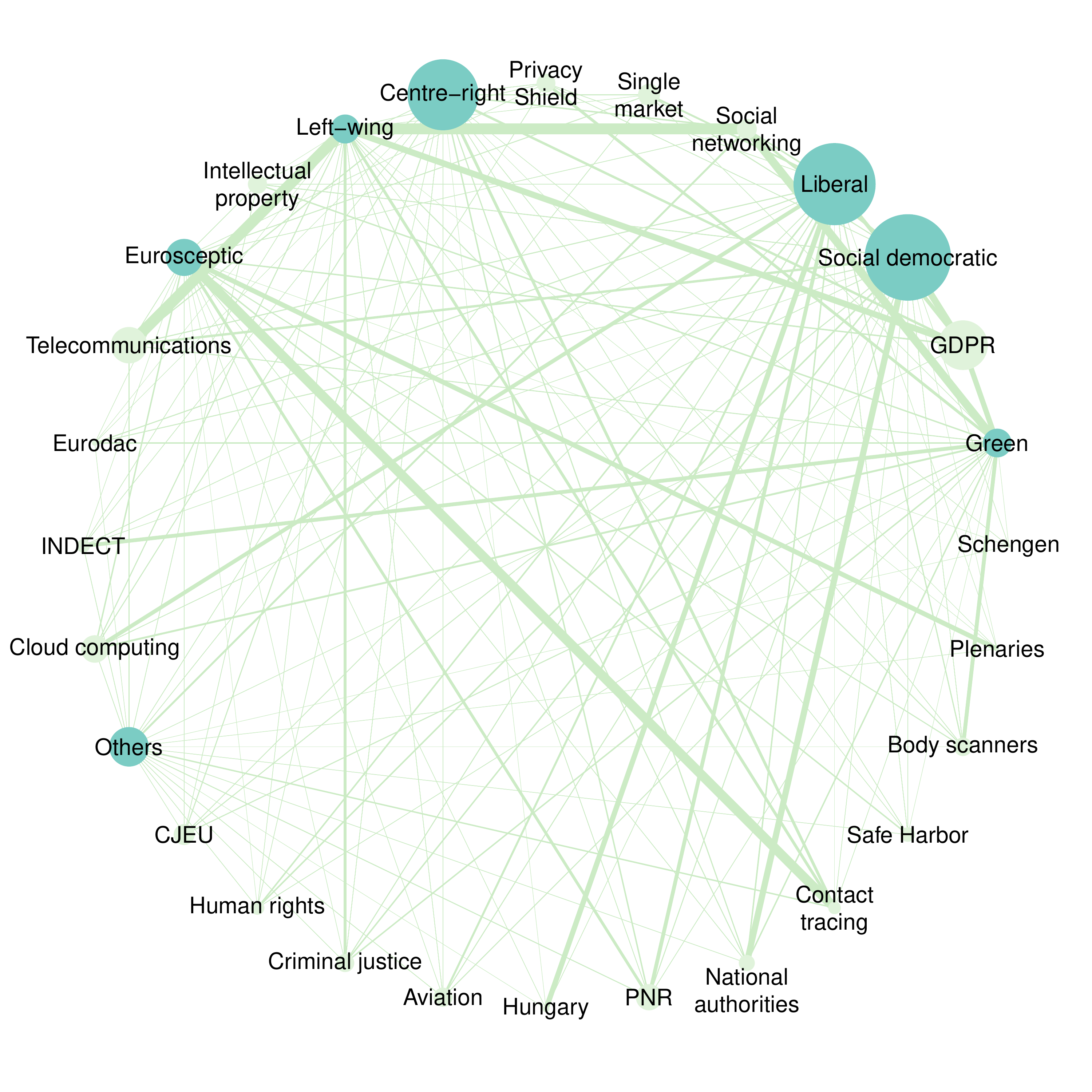}
\caption{A Network for Topics and Political Groups (the size of vertices is
  scaled by their degrees, while the width of edges is scaled according to the
  number questions asked about a given dominant topic)}
\label{fig: network}
\end{figure}

But to briefly continue the empirical analysis, Fig.~\ref{fig: network}
visualizes the noted network between the topics identified and the political
affiliations of the MEPs who have asked the questions related to these
topics. To recall: the edge weights and hence their widths in the visualization
refer to the times MEPs from a particular group asked questions about a given
topic, either alone or together with other MEPs. With this detail in mind, it
can be started by noting that the largest political group, the centre-right
group, is associated with many topics but none of these stand out in terms of
the volume of questions asked. In other words, it appears that the centre-right
group has not specialized into any given data protection topic. Instead, the
liberal group is associated with three topics in particular: cloud computing,
the rule of law issues in Hungary, and the passenger name record debacle. Then,
the social democratic group has been particularly active with questions related
to national authorities, including data protection authorities in the member
states, and the GDPR. The MEPs affiliated with the left-wing group, in turn,
have asked many questions related to social networking and social media, the
GDPR, and telecommunications. Also the green group has been active in debating
data protection issues in social networking, while the eurosceptic group has
been particularly keen to debate data protection questions related to contact
tracing. These observations align with those made earlier about the volume of
questions asked. The interests expressed toward data protection issues in social
networking by the left-wing and green groups and the eurosceptic group's
concerns about contact tracing help to explain the recent 2020s spikes in
Fig.~\ref{fig: groups}. That is, these topics are relatively new at least when
compared to the 2000s and 2010s debates on the PNR, data retention, and other
surveillance issues.

\section{Conclusion}\label{sec: conclusion}

This paper examined the largely uncovered terrain of data protection politics in
the European Union. The empirical material was based on the questions raised by
parliamentarians and the answers to these by the European Commissions in the
European Parliaments' plenaries from 1995 to early 2023. Given the lack of
existing research, the paper's research approach was exploratory, based on
quantitative text mining in general and topic modeling in particular. The
following four points summarize the results:

\begin{itemize}

\item{Data protection has been actively debated in the European Parliament
  plenaries throughout the past two decades. Although the volume of data
  protection questions raised by EU parliamentarians slightly increased in the
  mid-2000s during the GDPR's policy-making, stationarity characterizes all
  relevant time series; there are no notable ebbs and flows in these
  series. This result contradicts the argument that technology-related
  policy-making would be generally driven by some exceptional events
  \citep{Schneier20}. Instead, the time series results confirm the argument that
  the EU's overall data protection policy has shown strong path-dependency
  throughout the decades~\citep{Minkkinen19}. Data protection debates have also
  continued and even slightly increased in the Parliament's plenary sessions in
  the early 2020s.}

\item{In terms of references made to specific directives, regulations,
  decisions, and legislative proposals, the legacy data protection Directive
  95/46/EC leads the way, although the GDPR has stolen the references in more
  recent years. A similar result applies with communications and law proposals;
  the one for the GDPR has been the most referenced. Also the e-privacy
  Directive 2002/58/EC has frequently been referenced. Furthermore, Decision
  2008/997, which was repealed by Directive (EU) 2016/680, has received a lot of
  references, indicating that data protection questions related to law
  enforcement and criminal justice have also been actively debated by the
  parliamentarians. Typically, these references have occurred in the
  Commissions' replies to the questions rather than in the questions
  themselves.}

\item{Liberals, including those affiliated with the current Renew and its
  predecessors, have raised more data protection questions than other political
  groups. Social democratic, centre-right, green, left-wing, and eurosceptic
  groups have followed, in the order of listing. However, the volume of
  questions raised by liberals occurred particularly during the first part of
  the 2010s, whereas social democratic, green, left-wing, and eurosceptic groups
  have been active in more recent years. The volume of data protection questions
  raised by MEPs affiliated with the largest centre-right group, which includes
  the current EPP, has been large but without notable trends. Besides reflecting
  the sizes of these groups in the various EU parliaments over the years, the
  results implicitly hint that traditional political dimensions, such as the
  left-right axis, have not been a predominant force in the data protection
  debates.}

\item{Data protection questions have dealt with various different topics. The
  algorithmic computations conducted indicate that as many as sixty or even
  eighty distinct topics could be used to characterize the questions and the
  answers to these. Many distinct topics emerged even with the twenty-four
  topics preferred. Some of these are as could be expected: human rights and
  fundamental rights, the GDPR, the PNR, telecommunications (including its
  relation to the e-privacy directive), the Safe Harbor, the Privacy Shield, and
  the CJEU's decisions on transatlantic data flows and other big issues. There
  are also other topics that could be reasonably expected to emerge in
  parliamentary debates about data protection: law enforcement and criminal
  justice, the (digital) single market, national supervisory authorities, the
  Schengen area, social networking and social media, and cloud computing. Then,
  there are further topics for some unique issues that could be a little harder
  to anticipate beforehand: intellectual property, body scanners used in
  airports, Hungary, the INDECT research project, the Eurodac database, and
  contact tracing during the COVID-19 pandemic. All in all, these numerous
  different topics in the plenaries underline that European data protection
  politics go well-beyond the GDPR.}

\end{itemize}

These results confirm the prior expectation: data protection and privacy do seem
to provide political capital for European politicians, regardless of their
ideological positions and political affiliations. A good question for further
research would be whether and how the political data protection agenda-setting
by MEPs, whether via the plenaries or through other venues, actually affects the
legislative work in the EU. In this regard, there are various reforms underway
in the context of the envisioned digital single market. Some laws have already
been enacted. Further research is required to better understand how well data
protection issues have been addressed in these recent laws and
proposals. Regarding the recent questions raised by MEPs, it seems that the GDPR
has been the canonical answer given by the Commission of von der Leyen. As there
has been a continuous stream of data protection questions raised by politicians
(let alone by others), a question also remains about how well the GDPR can
curtail the new and emerging issues with new technologies. Finally, it would be
worthwhile to examine whether politicians can actually exploit their political
capital obtained through the political agenda-setting on data protection, that
is, whether the topic matters for citizens who elect the politicians.


\section*{Acknowledgements}

This research was funded by the Strategic Research Council at the Academy of Finland (grant number 327391).

\section*{Conflicts of Interest}

There is a conflict of interest with other researchers funded by the same grant (no. 327391) from the Strategic Research Council at the Academy of Finland.

\FloatBarrier
\section*{Appendix}

\begin{figure}[p!]
\centering
\includegraphics[width=\linewidth, height=18cm]{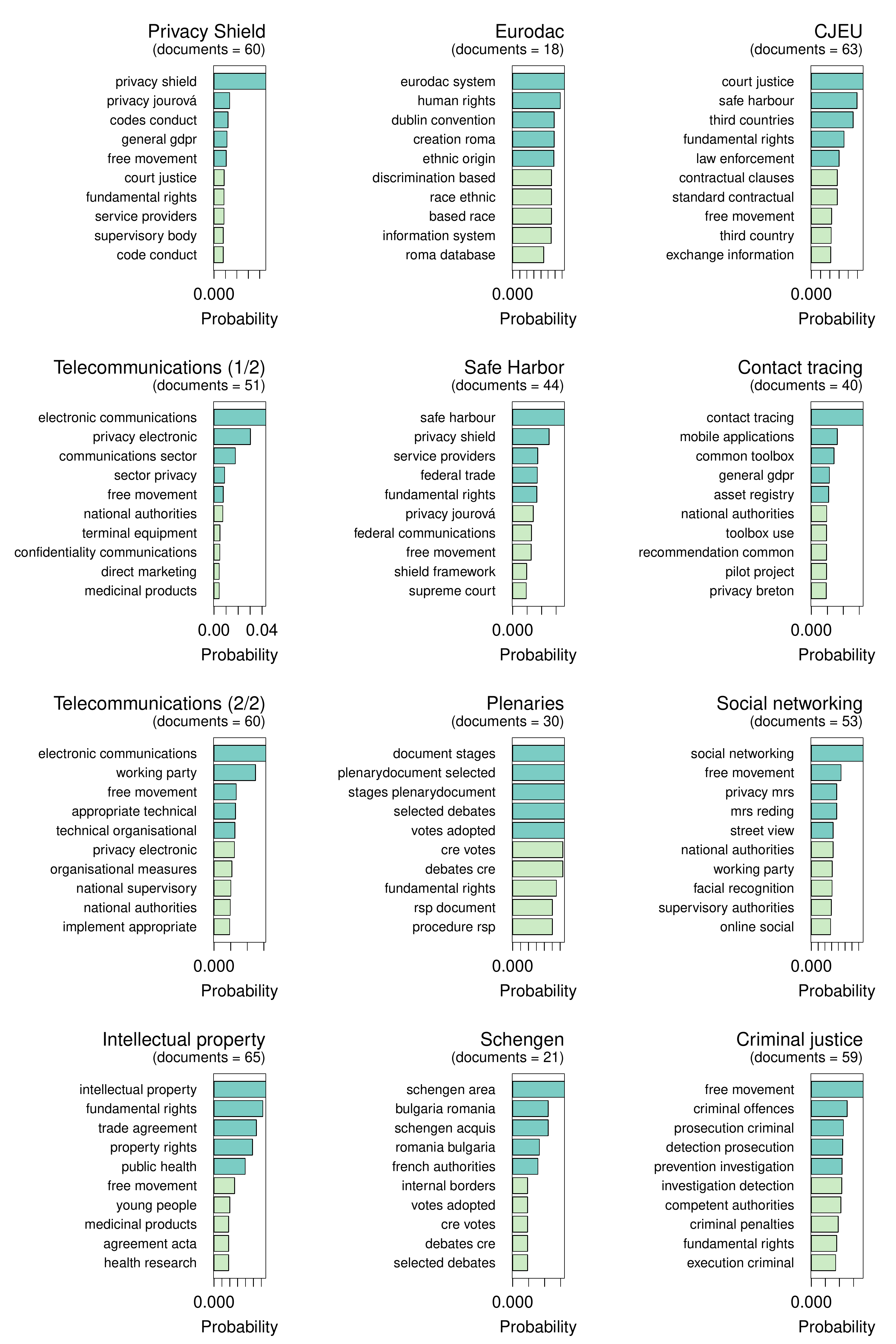}
\caption{Top-10 Bigrams for the final LDA Model (1/2)}
\label{fig: lda bigrams 1}
\end{figure}

\begin{figure}[p!]
\centering
\includegraphics[width=\linewidth, height=18cm]{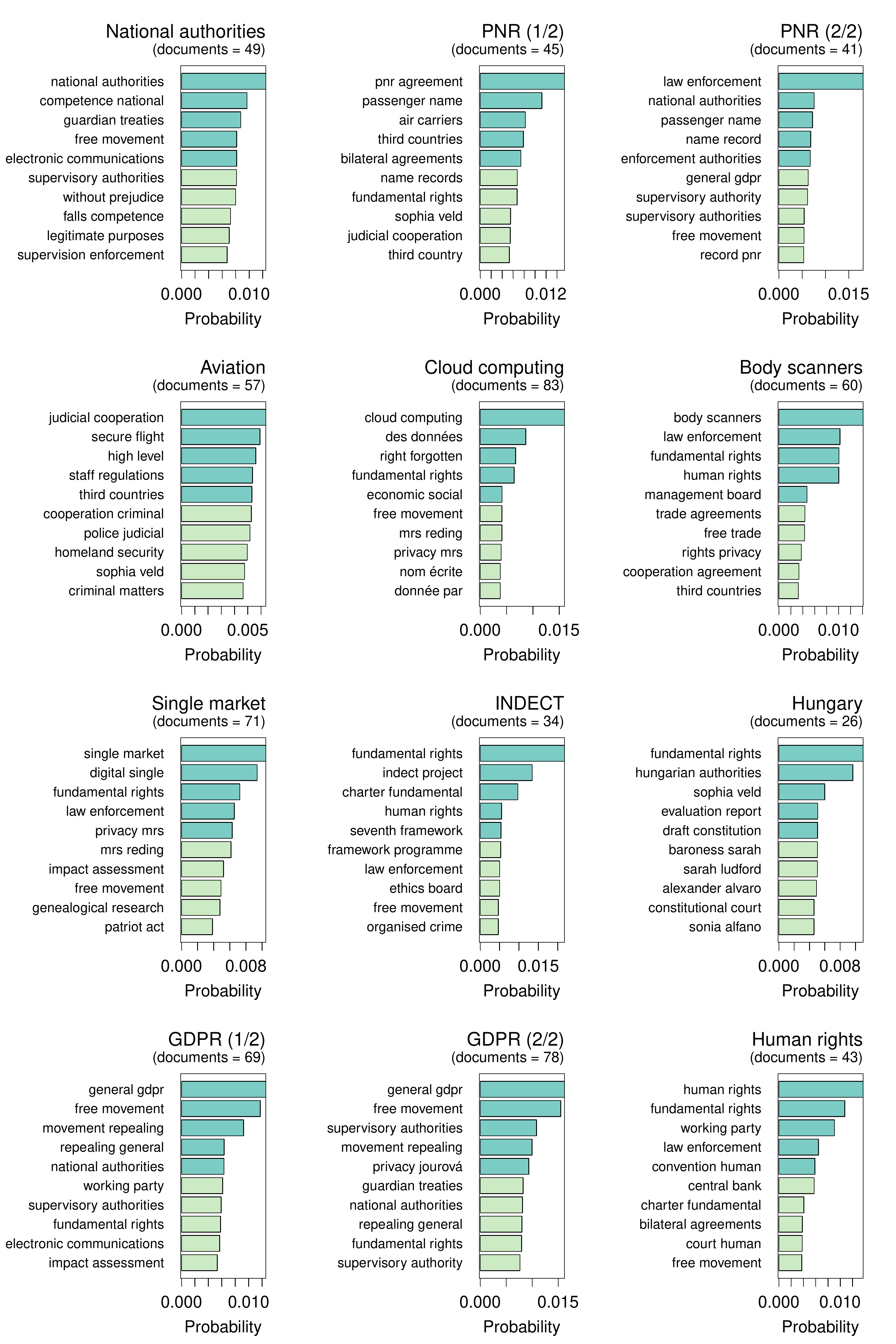}
\caption{Top-10 Bigrams for the final LDA Model (2/2)}
\label{fig: lda bigrams 2}
\end{figure}

\FloatBarrier

\balance
\bibliographystyle{apalike}

\end{document}